\documentclass[aps,prd,twocolumn,showpacs,floatfix,preprintnumbers,amsmath,amssymb,nofootinbibgrouped, notitlepage]{revtex4-1}

\input epsf
\usepackage{graphicx}
\usepackage{color}
\usepackage{mathtools}
\usepackage{mathbbol}

\newcommand{\beq}{\begin{equation}}
\newcommand{\eeq}{\end{equation}}
\newcommand{\barr}{\begin{eqnarray}}
\newcommand{\earr}{\end{eqnarray}}

\newcommand{\rme}{\textrm{e}}

\newcommand{\bs}{\boldsymbol}

\newcommand{\apjl}{Astrophys. J. Lett.}
\newcommand{\aap}{Astron. Astrophys.}

\newcommand{\mnras}{Mon. Not. R. Astron. Soc.}

\newcommand{\physrep}{Phys. Rep.}

\newcommand{\jcap}{J. Cosm. Astropart. Phys.}

\usepackage{color}
\newcommand{\chg}[1]{{#1}}

\newcommand{\lsim}{\mathrel{\hbox{\rlap{\lower.55ex\hbox{$\sim$}} \kern-.3em \raise.4ex \hbox{$<$}}}}
\newcommand{\gsim}{\mathrel{\hbox{\rlap{\lower.55ex\hbox{$\sim$}} \kern-.3em \raise.4ex \hbox{$>$}}}}

\begin{document}
\title{Correlation function of high-threshold \chg{regions} and \\
application to the initial \chg{small-scale} clustering of primordial black holes}
\author{Yacine Ali-Ha\"imoud}
\affiliation{Center for Cosmology and Particle Physics, Department of Physics,
New York University, New York, NY}
\date{\today}

\begin{abstract}
Primordial black holes (PBHs) have been brought back into the spotlight by LIGO's first direct detection of a binary-black-hole merger. One of the poorly understood properties of PBHs is how clustered they are at formation. It has important implications on the efficacy of their merging in the early Universe, as well as on observational constraints. In this work we study the initial clustering of PBHs formed from the gravitational collapse of large density fluctuations in the early Universe. We give a simple and general argument showing that, in this scenario, we do not expect clustering on very small scales beyond what is expected from a random, Poisson distribution. We illustrate this result explicitly in the case where the underlying density field is Gaussian. We moreover derive a new analytic expression for the two-point correlation function of large-threshold fluctuations, generalizing previous results to arbitrary separation, and with broader implications than the clustering of PBHs.

\end{abstract}

\maketitle

\section{Introduction}

The intriguing possibility that primordial black holes (PBHs) could have formed in the early Universe out of the collapse of rare, horizon-size, order-unity radiation fluctuations was first raised by Hawking \cite{Hawking_71}. Although more exotic formation scenarios have since then been suggested (see, e.g.~\cite{Sasaki_18} for a review), this remains the most studied to date. Hawking further posited that these ``collapsed objects [...] could stabilize clusters of galaxies, which, otherwise, appear mostly not to be gravitationally bound". While the nomenclature has changed since the early seventies, the question of the nature of dark matter remains as nagging now as it was then. Now more than ever, PBHs are an interesting dark matter candidate, as LIGO provides a new powerful way to search for them \cite{LIGO_05}, complementing the suite of observational tests that have already been proposed and/or carried out (see e.g.~\cite{Carr_10, Carr_16, Sasaki_18} for a review of constraints).

\begin{figure*}
\includegraphics[width =0.9\textwidth]{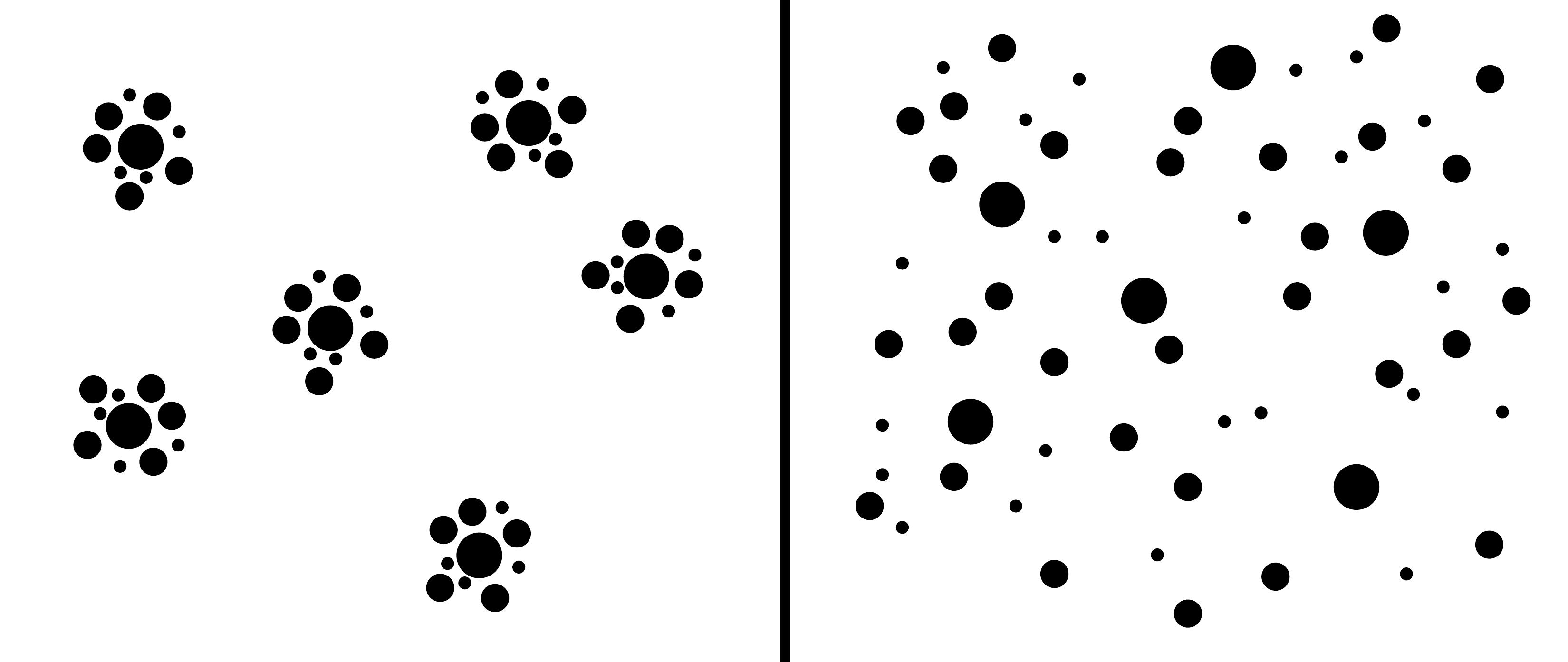}
\caption{Schematic representation of qualitatively different small-scale spatial distribution of PBHs at formation. On the left, PBHs are in dense clusters, as predicted in Ref.~\cite{Chisholm_06}. On the right, PBHs are distributed approximately randomly. In this work, we argue that the latter distribution is what is expected for PBHs forming from large density fluctuations. \chg{Note that this graphic is only schematic and ignores relativistic gauge issues.}}
\label{fig:scheme}
\end{figure*}

An important yet relatively poorly understood property of PBHs is their spatial clustering at formation. For one, if PBHs form in dense clusters \chg{(as in the left panel of Fig.~\ref{fig:scheme})}, they may quickly merge into larger black holes, and have a vastly different mass distribution at late times than they started with \cite{Chisholm_11, Clesse_15, GB_17}. In addition, observational implications of PBHs, hence constraints to their abundance \cite{Carr_10, Carr_16}, can be vastly different whether they are born clustered or mostly randomly distributed. In particular, it was argued \cite{GB_18} that clustered PBHs (or compact objects in general \cite{Metcalf_96}) could evade current microlensing constraints \cite{Macho_01, Eros_07}, as well as cosmic microwave background (CMB) limits \cite{Ricotti_08, YAH_17, Poulin_17} resulting from their accretion-powered energy injection \cite{Carr_81}. Last but not least, the merger rate of PBH binaries \cite{Nakamura_97, Sasaki_16, Bird_16, Raidal_17, YAH_17b} depends significantly on their initial small-scale clustering \cite{Clesse:2016vqa, Clesse_17}.

The first detailed study of the initial clustering of PBHs formed from the collapse of large fluctuations was undertaken in Ref.~\cite{Chisholm_06}. Assuming an underlying Gaussian density field, they computed the two-point correlation function of the PBH distribution, $\xi_{\rm pbh}(r)$. Using well-known analytic approximations, Ref.~\cite{Chisholm_06} found that in the limit of zero separation, $\xi_{\rm pbh}(0) \gg 1/P_1$, where $P_1$ is the probability to form a PBH in a horizon volume. This lead them to the conclusion that PBHs form in clusters, with a large mean occupation number $N_c \approx \xi_{\rm pbh}(0) P_1 \gg 1$. This interesting finding has not been revisited since then.

Here we argue on very general grounds, that in fact $1 + \xi_{\rm pbh}(0) = 1/P_1$, as one expects for objects randomly distributed on small scales. We moreover illustrate our general argument by studying the case of Gaussian perturbations, and point out the subtle point that mislead Ref.~\cite{Chisholm_06}. Along the way, we derive an analytic expression for the two-point correlation function of large-threshold fluctuations, Eq.~\eqref{eq:boxed}, generalizing existing results to arbitrary separation, and accurate even for moderately-large thresholds. This new result ought to be useful in more general setups, such as the study of biased tracers in large-scale structure \cite{Desjacques_18}.

\vspace{-8pt}

\section{Correlation function at zero lag}

We denote by $\delta$ the initial radiation density perturbation. The formation of a PBH at $\bs{r}$ takes place if the radiation field satisfies some criterion $\mathcal{C}[\delta]_{\bs{r}}$. A simple and often used approximation of this criterion is that the density perturbation smoothed over a horizon volume exceeds a critical value $\delta_c$, which depends on the shape of the fluctuation and the equation of state of the collapsing fluid. In reality the exact criterion $\mathcal{C}[\delta]$ is more complex \cite{Harada_15, Germani_18}. As we will see, our argument does not require knowing its explicit from. We denote the probability to form a PBH at position $\bs{r} = 0$ by
\beq
P_1 \equiv  P\left(\mathcal{C}[\delta]_0\right).
\eeq
The two-point correlation function of the PBH spatial distribution is the excess probability (over random) of finding two PBHs with separation $r$ \cite{Kaiser_84}
\barr
1 + \xi_{\rm pbh}(r) &=& \frac{P_2}{P_1^2} \equiv \frac{P\left(\mathcal{C}[\delta]_0 , \mathcal{C}[\delta]_{\bs{r}}\right)}{P_1^2} = \frac{P\left(\mathcal{C}[\delta]_{\bs{r}} \big{|}\mathcal{C}[\delta]_0\right)}{P_1}, ~~~\label{eq:xi-def}
\earr
where we have re-written the joint probability $P(\mathcal{C}_0,  \mathcal{C}_{\bs{r}})$ as the product of $P(\mathcal{C}_0) = P_1$ times the conditional probability $P(\mathcal{C}_{\bs{r}} | \mathcal{C}_0)$. The latter is always less than unity, and as a consequence, it must be that
\beq
1 + \xi_{\rm pbh}(r) \leq 1/P_1, \ \ \ \forall ~r.
\eeq
This inequality is saturated at zero separation, since $P(\mathcal{C}_0 | \mathcal{C}_0)  = 1$ (note that the correlation function need not be continuous at $r \rightarrow 0$ due to possible exclusion effects \cite{Mo_96, Baldauf_16, Desjacques_18}):
\beq
1 + \xi_{\rm pbh}(0) = 1/P_1. \label{eq:main}
\eeq
We emphasize that we did not make any specific assumption about the probability distribution of the underlying density field in this derivation; in particular, it applies whether the underlying field is Gaussian or not. We also stress that our argument is independent of the details of the formation criterion $\mathcal{C}[\delta]$.

Let us now explain how this implies that PBHs are \emph{initially} at most Poisson-clustered on small enough scales. The formation criterion should not depend on the density field much outside the horizon length at formation \cite{Young_14}, hence the PBH correlation function ought to drop rapidly at larger separations. The mean number density of PBHs is then approximately $P_1$ per correlation length cubed, i.e.~$n_{\rm pbh} \sim P_1/V_H$, where $V_H$ is the horizon volume (this supposes that one PBH is formed per horizon volume if the criterion is satisfied). Therefore, $\xi_{\rm pbh}(r)$ is bounded by a function whose value at the origin is approximately $1/(n_{\rm pbh} V_H)$, and which quickly drops at separations greater than a horizon size. This bounding function is approximately $\delta_{\rm Dirac}(\bs{r})/n_{\rm pbh}$, smoothed over a horizon volume, which what is expected for a Poisson distribution of finite-size objects. Note that the clustering can in fact be sub-Poissonian at small separations due to exclusion effects \cite{Mo_96, Baldauf_16, Desjacques_18}. We expect such effects to matter only at separations of the order of a few horizon lengths, much smaller than scales relevant to any observational tests of PBHs. \chg{We also emphasize that this discussion can only be made fully quantitative with a rigorous relativistic treatment, outside our scope.}

\vspace{-10pt}

\section{Correlation function of rare overdensities of a Gaussian field} \label{sec:corr-func}

Let us now specify to the case where $\delta$ is a Gaussian random field, whose statistics are hence entirely determined by its two-point correlation function $\xi(r) \equiv \langle \delta(0) \delta(\bs{r}) \rangle \equiv \sigma^2 w(r)$, where $\sigma^2 \equiv \xi(0) \equiv \langle \delta^2 \rangle$ is the variance, and $0 \leq w(r) \leq 1$. The normalized correlation $w(r)$ approaches unity for small separations, and zero for large separations. We consider the clustering of objects with the simple formation criterion $\delta > \delta_c$, and denote by $\nu \equiv \delta_c/\sigma$ the formation threshold in units of the standard deviation. We will focus in particular on the case $\nu \gg 1$, which is typically expected if PBHs are to form out of the rare order-unity fluctuations of an otherwise nearly smooth background. The probability of being above threshold is
\beq
P_1 = \frac12 \textrm{erfc}\left(\frac{\nu}{\sqrt{2}}\right), \label{eq:P1-Gaussian}
\eeq
and the probability that two regions separated by $\bs{r}$ are both above threshold is \cite{Kaiser_84} 
\barr
P_2 &=&  \int_{\nu}^\infty \frac{d x_1}{\sqrt{2 \pi}} \int_{\nu}^\infty \frac{d x_2}{\sqrt{2 \pi}} \frac1{\sqrt{1- w^2}} \nonumber\\
&&\times \exp\left[ - \frac{x_1^2 + x_2^2 - 2 w x_1 x_2}{2(1 - w^2)} \right].  
\earr
We now rewrite this integral in a more convenient way. We start by changing variables to $x_{\pm} \equiv (x_2 \pm x_1)/\sqrt{2}$, which are two uncorrelated, Gaussian-distributed variables, as can be seen when rewriting the exponent as
\beq
\frac{x_1^2 + x_2^2 - 2 w x_1 x_2}{1 - w^2} = \frac{x_-^2}{1 -w} + \frac{x_+^2}{1 + w}.
\eeq
The integration domain $x_1 > \nu, x_2 > \nu$ corresponds to $x_- \in (-\infty, \infty)$, $x_+ > \sqrt{2}~ \nu + |x_-|$. We therefore get
\barr
P_2 &=&  \frac1{2 \pi} \int_{-\infty}^\infty \frac{d x_-}{\sqrt{1 - w}} \exp\left[ - \frac{x_-^2}{2(1-w)} \right] \nonumber\\
&&\times \int_{\sqrt{2} \nu + |x_-|}^{\infty} \frac{d x_+}{\sqrt{1 + w}} \exp\left[ - \frac{x_+^2}{2(1+w)} \right].
\earr
The innermost integral can be expressed in terms of a complementary error function. A final change of variables to $x = x_-/\sqrt{1-w}$ leads to the following form, well suited for numerical evaluation:
\barr
P_2 &=& \sqrt{\frac2{\pi}} \int_{0}^\infty dx~ \rme^{-x^2/2} \nonumber\\
&&\times \frac12 \textrm{erfc}\left[ \frac{\nu}{\sqrt{1+w}} \left( 1 +  \sqrt{\frac{1-w}2} \frac{x}{\nu} \right) \right]. \label{eq:P2-exact}
\earr
So far this expression is exact, and holds for arbitrary $\nu$. Let us now consider the case where $\nu \gg 1$. We recall that for large argument, the complementary error function can be approximated by 
\beq
\textrm{erfc}(X) = \frac{\rme^{-X^2}}{\sqrt{\pi}~X} \left[ 1 + \mathcal{O}(1/X) \right], \ \ \ X \gg 1. \label{eq:erfc}
\eeq
For $\nu \gg 1$, we may use this asymptotic expression for $P_1$ as well as the erfc inside Eq.~\eqref{eq:P2-exact} for any value of $x \geq 0$ and $0 \leq w \leq 1$. We then find the following asymptotic expression for the ratio $P_2/P_1$, in the large-$\nu$ limit:
\barr
\frac{P_2}{P_1} &\approx& \sqrt{\frac{1+w}{\pi}}  \int_{0}^\infty dx ~ F(x)~ \rme^{- S(x)}, \label{eq:P2_ov_P1}\\
F(x) &\equiv& \left( 1 +  \sqrt{\frac{1-w}2} \frac{x}{\nu} \right)^{-1}, \\ 
S(x) &\equiv&  \frac{x^2 - \nu^2}2 + \frac{(\sqrt{2} \nu + x \sqrt{1-w})^2 }{2(1 + w)} \nonumber\\
&=&  \frac1{1 + w} \left( x + \frac{\nu}{\sqrt{2}}\sqrt{1 - w} \right)^2.
\earr
The contributions of $x \gtrsim 1$ are exponentially suppressed, and we may therefore approximate the prefactor $F(x) \approx 1$, while keeping the full expression for the exponent $S(x)$. The integral over $x$ can then be computed analytically, giving
\beq
\frac{P_2}{P_1} \approx \frac{1 + w}{2} \textrm{erfc}\left[ \sqrt{\frac{1 - w}{1 + w}} \frac{\nu}{\sqrt{2}}\right],  \ \ \nu \gg 1.
\eeq 
Dividing by $P_1$ and using Eqs.~\eqref{eq:xi-def} and \eqref{eq:P1-Gaussian}, we arrive at our main new result, \emph{valid for any} $w \in [0, 1]$:
\barr
\boxed{1 + \xi_{\nu}(r) \approx (1 + w)   \frac{\textrm{erfc}\left( \sqrt{\frac{1 - w}{1 + w}}~ \nu/\sqrt{2}\right)}{\textrm{erfc}(\nu/\sqrt{2})}, \ \ \ \nu \gg 1.~~~~~~~~~ \label{eq:boxed}} 
\earr
This expression is the asymptotic form of the two-point correlation function of a thresholded process, in the limit of large threshold, but arbitrary separation. We cannot further expand the numerator without making additional assumptions about the relative magnitude of $\nu$ and $1/\sqrt{1 - w}$. We also note that, although one could consistently expand the denominator in the large-$\nu$ limit, the expression we have adopted is more accurate when $w \rightarrow 0$, for large but finite $\nu$.

We now consider limiting cases for $w$. First, for $r \rightarrow 0$, hence $w \rightarrow 1$, we find, for any fixed $\nu \gg 1$,
\barr
P_2 &\approx&  \left( 1 - \sqrt{1 - w^2} \frac{\nu}{\sqrt{2 \pi}}\right) P_1 \nonumber\\
&\approx& P_1 - \frac{\sqrt{1 - w^2} }{2 \pi} \rme^{- \nu^2/2},
\earr
where in the second equality we have expanded $P_1$ in the large-$\nu$ limit. This expression matches Equation (10) of Jensen and Szalay \cite{Jensen_86}. In particular, we see that $P_2 \rightarrow P_1$ for $w \rightarrow 1$, i.e.~$r \rightarrow 0$, and we recover Eq.~\eqref{eq:main}.

Now, for $\nu \gg 1/\sqrt{1 - w}$, we may expand the complementary error function in Eq.~\eqref{eq:boxed}, and obtain
\beq
1 + \xi_{\nu}(r) \approx \frac{(1 + w)^{3/2}}{(1 - w)^{1/2}} ~\rme^{\frac{w}{1 + w} \nu^2} , \ \ \ \nu \gg 1/\sqrt{1 - w}.
\eeq
For large separations, $w \rightarrow 0$, and $w/(1 + w) = w + \mathcal{O}(w^2)$. Provided $\nu^2 w^2 \ll 1$, we may neglect the term of order $w^2 \nu^2$, and obtain the following result, derived by Politzer and Wise \cite{Politzer_84}:
\beq
1 + \xi_{\nu}(r) \approx  \rme^{w \nu^2} , \ \ \ w \ll 1/\nu \ll 1. \label{eq:approx}
\eeq
Finally, if the condition $ w \ll 1/\nu^2 \ll 1$ is satisfied, we recover Kaiser's well-known result \cite{Kaiser_84}, $\xi_{\rm pbh}(r) \approx \nu^2 w(r)$.

\begin{figure*}
\includegraphics[width =0.9\textwidth]{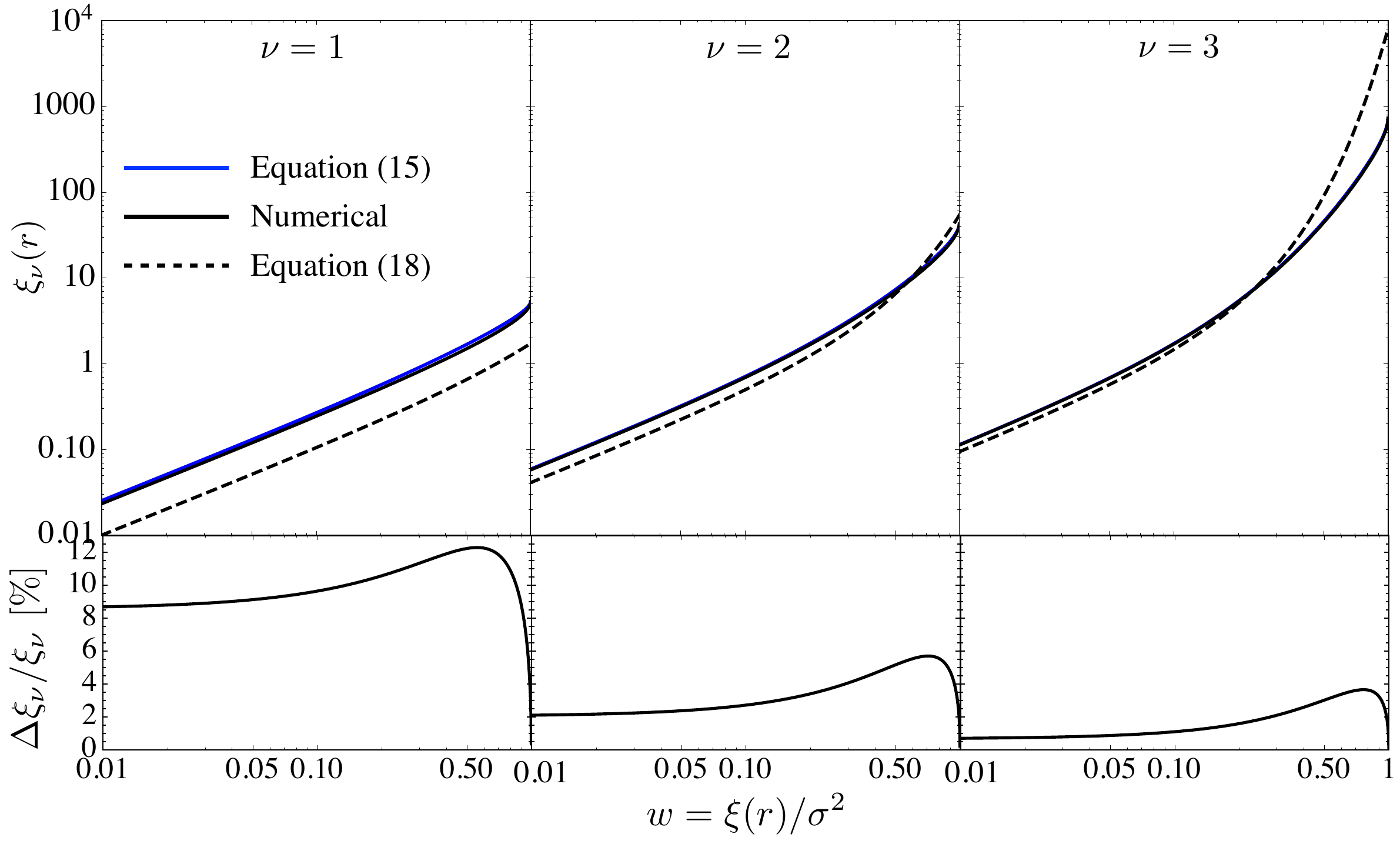}
\caption{Two-point correlation function of regions with density $\delta > \nu \sigma$, using the exact numerical expression (solid black), our analytic expression \eqref{eq:boxed} valid in the large-$\nu$ regime, but arbitrary $w$ (solid blue), and using the small-$w$ approximation, Eq.~\eqref{eq:approx} (dashed black). The solid black and solid blue lines are nearly undistinguishable except for $\nu = 1$. The lower panels show the small fractional error (in percent) between our simple analytic approximation \eqref{eq:boxed} and the numerical correlation function.}
\label{fig:corr}
\end{figure*}

We compared our analytic approximation \eqref{eq:boxed} to the exact correlation function obtained from numerically integrating Eq.~\eqref{eq:P2-exact}, and found excellent agreement for all $0 \leq w \leq 1$, and for large $\nu$. The approximation is good even for $\nu \sim 1$: we find a maximum relative error on $\xi_{\nu}$ of $4, 6$ and 12 \% for $\nu = 3, 2$ and 1, respectively (see bottom panel of Fig.~\ref{fig:corr}). Should more accurate analytic approximations be needed, one could easily continue our expansion to higher orders in $1/\nu$, by first expanding the error function to the next order, and consistently keeping track of prefactors in Eq.~\eqref{eq:P2_ov_P1}.

As already pointed out by Jensen and Szalay \cite{Jensen_86}, the approximation \eqref{eq:approx} noticeably over-estimates the correct result at $w \rightarrow 1$, i.e.~small separation. We illustrate this in Fig.~\ref{fig:corr} (similar to Fig.~1 of \cite{Jensen_86}), where we show $\xi_{\nu}$ computed numerically, alongside our new result \eqref{eq:boxed} and the small-$w$ approximation \eqref{eq:approx}. 

It was Equation \eqref{eq:approx} that was used in the derivation of Ref.~\cite{Chisholm_06}, where it was extrapolated to $r \rightarrow 0$, hence $w \rightarrow 1$, where it does not hold. This mislead to the conclusion that PBHs form in clusters, as opposed to being Poisson-distributed on small enough scales.

\vspace{-8pt}
\section{Conclusions}

We have argued on very general grounds that PBHs are not expected to form in clusters \chg{(depicted in the left panel of Fig.~\ref{fig:scheme})}, at least if they result from the collapse of horizon-size, order-unity density fluctuations. We do not consider other, more exotic formation mechanisms, see e.g.~\cite{Belotsky_18}. We illustrated our general derivation by studying the case of underlying Gaussian perturbations. We derived a new analytic approximation for the correlation function of large-threshold fluctuations, valid for arbitrary separations. Our derivation is rather simple and can easily be extended to higher order in $1/\nu$. 

\chg{Specifically, we showed that PBHs are not born clustered \emph{beyond Poisson} on small scales. They are \emph{still initially clustered}, in the sense that they have a non-vanishing two-point correlation function (2pcf). We do not attempt to estimate this initial 2pcf in this work, and our calculation in Section \ref{sec:corr-func} should be understood as a toy model. Indeed, such a calculation would require, first, a detailed relativistic criterion for the formation of PBHs, a topic which is still under investigation \cite{Germani_18}. Secondly, computing the 2pcf on super-horizon scales at the time of PBH formation necessitates a thorough discussion of gauge issues. We refer the reader to Refs.~\cite{Young_15, Tada_15} for studies of the initial 2pcf of PBHs in the presence of primordial non-Gaussianity. }

Let us also emphasize that our work is focused on the \emph{initial} clustering of PBHs. Just like for any non-relativistic collisionless matter, \chg{PBH density fluctuations will grow, first linearly,} and eventually form clusters and non-linear structures \cite{Meszaros_75, Carr_77}. \chg{The 2pcf of PBHs around $z \sim 10^4-10^5$ is relevant to the computation of the PBH binary merger rate \cite{Nakamura_97, Raidal_17, YAH_17b}, and we point to Refs.~\cite{Desjacques_18b, Ballesteros_18} for recent attempts to estimate its effect.} The late-time clustering could affect observational implications of PBHs, such as their impact on CMB anisotropies due to non-linear motions \cite{YAH_17, Poulin_17}, or the evolution of PBH binaries formed in the early Universe \cite{YAH_17b}. \chg{We do not attempt to study any of these observational consequences in the present work.}

In addition, Poisson fluctuations due to the discrete nature of PBHs have been put forward as possible seeds of cosmic structure \cite{Carr_83, Carr_18}, \chg{and as a possible explanation of the cosmic infrared background \cite{Kashlinsky:2016sdv}}. Testing these interesting proposals, and the PBH hypothesis in general, will be made considerably simpler now that the question of the initial clustering of PBHs has been clarified. 

Our calculation ought to be valuable beyond the study of PBHs, as it clarifies a point often misunderstood in the literature on large-scale structure, and that can be obscured when considering higher-order statistics (see e.g.~\cite{Franciolini_18} and references therein). Modern analytic approaches are of course more sophisticated than the simple threshold criterion that we studied \cite{BBKS, Desjacques_18}. However, our uncovering of a new, extremely simple analytic approximation, that was missed in three decades of research, suggests that our method might also be fruitful for the study of more realistic halo formation criteria.

\vspace{-18pt}
\subsection*{Acknowledgements}

I thank Marc Kamionkowski, Ely Kovetz, Chiara Mingarelli, Fabian Schmidt, Roman Scoccimarro, and Matias Zaldarriaga for timely and thorough feedback on this manuscript. I am grateful to Sandrine Codis for pointing out the possibility of an exclusion region at small separation. I thank Pasquale Serpico for comments that helped me clarify the precise scope of this work.

\bibliography{pbh_gw.bib}

\end{document}